# Model Transformations in Practice Workshop

Jean Bézivin, Bernhard Rumpe, Andy Schürr, and Laurence Tratt

University of Nantes, TU Darmstadt, TU Braunschweig, King's College London
`http://sosym.dcs.kcl.ac.uk/events/mtip/`

## 1 Background

Model Transformations in Practice (MTiP) 2005 was a workshop which provided a forum for the model transformation community to discuss practical model transformation issues. Although many different model transformation approaches have been proposed and explored in recent years, there has been little work on comparing and contrasting various approaches. Without such comparisons, it is hard to assess new model transformation approaches such as the upcoming OMG MOF/QVT recommendation, or to discern sensible future paths for the area. Our aims with the workshop were to create a forum that would help lead to an increased understanding of the relative merits of different model transformation techniques and approaches. A more advanced understanding of such merits is of considerable benefit to both the model transformation and wider modelling communities.

## 2 Workshop Format

In order to achieve the workshops' aims, we took an unusual approach in the Call for Papers (CfP). We decided that the workshop would focus on underlying model transformations mechanisms, concepts, languages and tools, development environments, libraries, practises and patterns, verification and optimization techniques, traceability and composeability issues, applicability scope, deployment techniques, and so on. In order to achieve aim, we detailed a specific mandatory example that all submissions had to tackle (detailed in section 5), in order that it would be easier to compare and contrast submissions. Authors were asked to take a particular model transformation approach and structure their submission as follows:

1. An overview of the authors' chosen model transformation approach.
2. The required aspects of the mandatory model transformation example.
3. Optionally, additional aspects of the mandatory model transformation example.
4. Optionally, extra model transformations chosen by the authors from a list of alternatives.
5. Results and discussion.

Authors were asked to consider and discuss, where relevant, the following issues with regard to their chosen approach:

- Composition of transformations.
- Robustness and error handling,
- Debugging support.
- Flexibility, overall usability and power of the chosen approach.
- Whether the approach can express bidirectional and / or incremental (sometimes known as change propagating) transformations.
- Technical aspects such as the ability to deal with model exchange formats, modelling tool APIs, and layout updates.

## 3 Accepted Submissions

Because of the unusual demands of our CfP, we were pleasantly surprised at both the quantity and quality of submissions. In the end we accepted the following eight submissions:

Model Transformation by Graph Transformation: A Comparative Study
*Gabriele Taentzer, Karsten Ehrig, Esther Guerra, Juan de Lara, Laszlo Lengyel, Tihamer Levendovszky, Ulrike Prange, Daniel Varro, Szilvia Varro-Gyapay*, Technische Universität Berlin, Universidad Carlos III de Madrid, Universidad Autonoma de Madrid, Budapest University of Technology and Economics

Model Transformation with Triple Graph Grammars
*Alexander Königs*, University of Technology Darmstadt

Kent Model Transformation Language
*D.H.Akehurst, W.G.Howells, K.D.McDonald-Maier*, University of Kent

Practical Declarative Model Transformation With Tefkat
*Michael Lawley, Jim Steel*, DSTC, University of Rennes

Transforming Models with ATL
*Frédéric Jouault, Ivan Kurtev*, INRIA

Model Transformation Approach Based on MOLA
*Audris Kalnins, Edgars Celms, Agris Sostaks*, University of Latvia

On Executable Meta-Languages applied to Model Transformations
*Pierre-Alain Muller, Franck Fleurey, Didier Vojtisek, Zoé Drey, Damien Pollet, Frédéric Fondement, Philippe Studer, Jean-Marc Jézéquel*, IRISA/INRIA, France, EPFL/IC/UP-LGL, INJ, Switzerland, Université de Haute-Alsace

Model Transformation in Practice Using the BOC Model Transformer
*Marion Murzek, Gerti Kappel, Gerhard Kramler*, Vienna University of Technology

With so many high quality submissions to pick from, choosing only two for inclusion in these proceedings was an inevitably difficult task. However we believe that the two papers that the programme committee voted to select are indicative of the overall high quality of submissions.

## 4   Programme Committee

The workshop had a programme committee which reflected many of the different parts of the model transformation community. The programme committee performed sterling work in reviewing the CfP, voting on papers to accept and so on. The programme committee consists of:

| | |
|---|---|
| Wim Bast | Compuware, Netherlands |
| Tony Clark | Xactium, UK |
| Krzysztof Czarnecki | University of Waterloo, Canada |
| Gregor Engels | University of Paderborn, Germany |
| Kerry Raymond | DSTC, Australia |
| Robert France | Colorado State University, USA |
| Jens Jahnke | University of Victoria, Canada |
| Jean-Marc Jézéquel | University of Rennes, INRIA, France |
| Stuart Kent | Microsoft, UK |
| Gabor Karsai | Vanderbilt University, Tennessee, USA |
| Gregor Kiczales | University of British Columbia, Canada |
| Reiko Heckel | University of Leicester, UK |
| Dániel Varró | Budapest University of Technology and Economics, Hungary |
| R. Venkatesh | Tata Consultancy Services, India |
| Albert Zündorf | University of Kassel, Germany |

## 5   Mandatory Example

All submissions were asked to tackle the example as outlined in this section. The example itself is a slight variation on the well known 'class to RDBMS' transformation. This example was chosen because, despite its relative simplicity, it tends to exercise a broad class of model transformation features. Perhaps inevitably after the release of this example, prospective authors found small ambiguities, missing details, and even the odd small mistake in the specification. We kept the workshop website up to date with 'errata' on the CfP, and informally suggested to authors that in the event of doubt on their part, they were welcome to choose a particular path provided they documented it appropriately.

The rest of this section contains the model transformation specification as it was defined in the CfP which the reader will find useful when reading the two papers selected from the MTiP workshop.

### 5.1   Meta-models

The meta-model for class models is shown in figure 1. The following OCL constraint is also part of the model (the `allAttributes` operation returns a class's local and inherited attributes):

```
context Class inv:
    allAttributes()->size > 0 and
    allAttributes()->exists(attr | attr.is_primary = true)
```

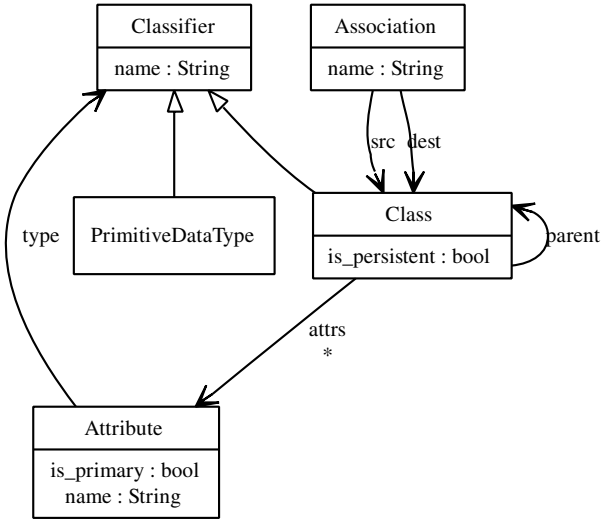

**Fig. 1.** Class meta-model

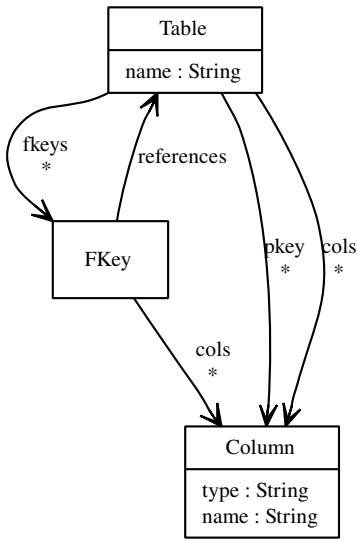

**Fig. 2.** RDBMS meta-model

A model consists of classes and directed associations. A class consists, possibly via inheritance, of one or more attributes, at least one of which must be marked as constituting the classes' primary key. An attribute type is either that of another user class, or of a primitive data type (e.g. String, Int). Associations are considered to have a `1` multiplicity on their destination. Submissions may assume the presence of standard data-types as instances of the `PrimitiveDataType` class.

The meta-model for RDBMS models is shown in figure 2. An RDBMS model consists of one or more tables. A table consists of one or more columns. One or more of these columns will be included in the `pkey` slot, denoting that the column forms part of the tables primary key slot. A table may also contain zero or more foreign keys. Each foreign key refers to the particular table it identifies, and denotes one or more columns in the table as being part of the foreign key.

**Transformation.** This version of the transformation contains several subtleties that authors will need to be aware of. In order to facilitate comparisons between approaches, authors should ensure that they accurately implement the transformation.

1. Classes that are marked as persistent in the source model should be transformed into a single table of the same name in the target model. The resultant table should contain one or more columns for every attribute in the class, and one or more columns for every association for which the class is marked as being the source. Attributes should be transformed as per rules 3 – 5.
2. Classes that are marked as non-persistent should not be transformed at the top level. For each attribute whose type is a non-persistent class, or for each association whose `dst` is such a class, each of the classes' attributes should be transformed as per rule 3. The columns should be named *name_transformed attr* where *name* is the name of the attribute or association in question, and *transformed attr* is a transformed attribute, the two being separated by an underscore character. The columns will be placed in tables created from persistent classes.
3. Attributes whose type is a primitive data type (e.g. String, Int) should be transformed to a single column whose type is the same as the primitive data type.
4. Attributes whose type is a persistent class should be transformed to one or more columns, which should be created from the persistent classes' primary key attributes. The columns should be named *name_transformed attr* where *name* is the attributes' name. The resultant columns should be marked as constituting a foreign key; the `FKey` element created should refer to the table created from the persistent class.
5. Attributes whose type is a non-persistent class should be transformed to one or more columns, as per rule 2. Note that the primary keys and foreign keys of the translated non-persistent class need to be merged in appropriately, taking into consideration that the translated non-persistent class may contain primary and foreign keys from an arbitrary number of other translated classes.
6. When transforming a class, all attributes of its parent classes (which must be recursively calculated), and all associations which have such classes as a `src`, should be considered. Attributes in subclasses with the same name as an attribute in a parent class are considered to override the parent attribute.

7. In inheritance hierarchies, only the top-most parent class should be converted into a table; the resultant table should however contain the merged columns from all of its subclasses.

Notes on the transformation:

- Rules 2, 4 and 5 are recursive – the 'drilling down' into attributes' types can occur to an arbitrary level.
- Associations do not directly transform into elements; however each association which has a particular class as a `src` must be considered when transforming that class into a table and / or columns.
- When merging the transformation of a non-persistent class, care must be taken to handle the primary and foreign keys of the transformed class appropriately.
- Foreign keys, primary keys and so on should point to the correct model elements – transformations which create duplicate elements with the same names are not considered to provide an adequate solution.

Authors are encouraged to take particular note of the following points when they create their transformations:

- The recursive nature of the drilling down.
- The creation of foreign keys.
- Associations.

**Example Execution.** Figures 3 and 4 show the example input and output to the class to RDBMS transformation example.

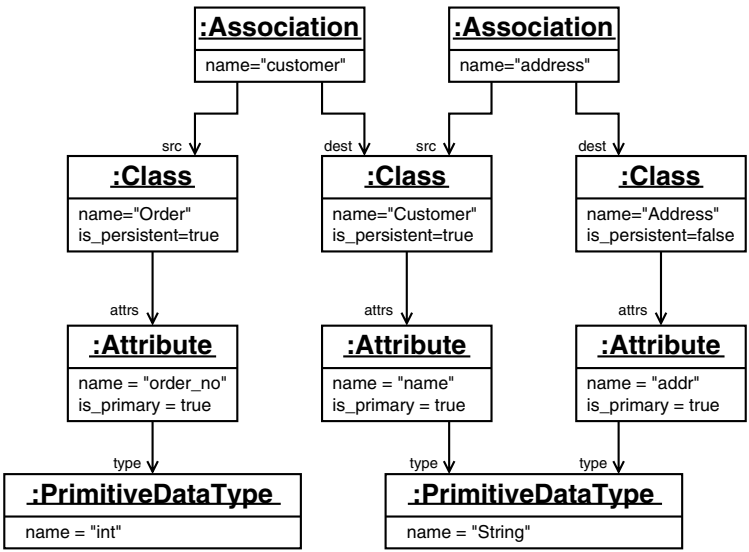

**Fig. 3.** Example input

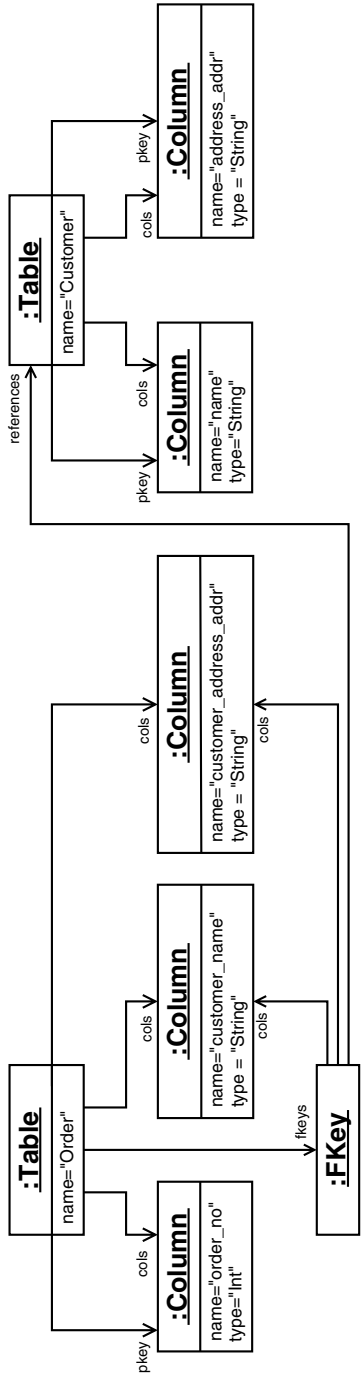

**Fig. 4.** Example output

## 6   Workshop Outcomes

The workshop itself was a lively, and well attended affair. We devoted a substantial portion of the day to discussion. Much of this related to the model transformation approaches presented, and their relation to other approaches not presented (e.g. the forthcoming QVT standard). In no particular order, some of the points raised during discussion were as follows:

- Current model transformation approaches lack scalability in two aspects: their efficiency, and their code organization. The latter would be aided by features such as modularity.
- The relationship of model transformations to normal compilers could fruitfully be explored.
- A lack of formalization of model transformation approaches, and consequent inability to reason reliably about model transformations.
- Are bidirectional transformations practical and / or desirable?
- The importance of tracing information for tool users to track their transformations.
- Difficulties in making diagrammatic syntaxes for all aspects of model transformations.
- A need for more sophisticated taxonomies of model transformation systems.
- A need to define the relationship of semantics preserving model transformations to the concept of refinement.

## 7   And Finally...

We would like to thank the authors of papers, the programme committee, and all those who turned up and participated on the day itself for making the MTiP workshop a success. Due to the interest in this subject, we anticipate holding another workshop on this subject to which you are all cordially invited!